# Controllable Growth and Characterization of α- and β-phase MnSe by Chemical Vapor Deposition

Jennifer E. DeMell[1,2,*], Elias Kallon[2], Michael Pedowitz[2], Jimmy C. Kotsakidis[2], Ihteyaz Aqaeed Avash[2], Kevin M. Daniels[2]

[1] Laboratory for Physical Sciences, College Park, MD, 20742
[2] Institute for Research and Applied Physics, University of Maryland, College Park, MD

*Corresponding Author: jdemell@lps.umd.edu

## Abstract

Manganese selenide (MnSe) is a promising air-stable two-dimensional magnetic semiconductor, for which theory predicts robust ferromagnetism in monolayers with Curie temperatures approaching 250 K. However, the crystallographic phases and magnetic properties of thin-film MnSe grown by scalable methods remain poorly understood. Here, we demonstrate controllable growth of α- and β-phase MnSe on C-face sapphire using a three-zone chemical vapor deposition process with elemental Se and $MnCl_2$ precursors in an Ar/$H_2$ atmosphere. Using Raman spectroscopy, atomic force microscopy, scanning electron microscopy, and X-ray photoelectron spectroscopy, we show that our process yields phase-pure α-MnSe nanorods and β-MnSe triangular flakes with lateral sizes up to 20 μm and thicknesses of 15–30 nm. Low-temperature photoluminescence of the β-phase films reveals a band gap of ≈ 2.0 eV. Systematic variation of growth parameters shows that precursor vapor pressure, rather than $H_2$ partial pressure, is the dominant factor controlling lateral flake size. Vibrating-sample magnetometry measurements reveal a Néel temperature of 53 K in the β-phase films, providing clear evidence of antiferromagnetism in the multilayer regime and establishing MnSe as a tunable platform for 2D spintronic and optoelectronic devices.

# Introduction

Two-dimensional materials such as graphene and transition metal chalcogenides hold significant promise for advancing low-dimensional memory and logic. For instance, transition metal dichalcogenides (e.g., $WS_2$) and monochalcogenides (e.g., FeSe) have been found to exhibit a wide range of physical phenomena including superconductivity, valley and exciton optical effects, and non-trivial topological phases.[1–4] Similarly, MnSe has shown ferroelectric behavior and pressure-induced superconductivity,[2] yet has received considerably less attention than its mono- and di-chalcogenide cousins due to its difficulty to reliably synthesize into large 2D sheets. This difficulty partly arises because low-dimensional MnSe can exist in four phases: α (rock salt), β (zinc blende), γ (wurtzite), and T ((111) face of the α phase), with each phase exhibiting a distinct magnetic behavior. The most widely studied of these phases is the α phase, which typically grows in the form of air-stable antiferromagnetic ribbons (Néel temperature, $T_N$ = 150 K) with a reported band gap of 3.0 eV.[5–7] The β-phase has been previously reported as ferromagnetic (Curie temperature, $T_C$ = 42.3 K), though the band gap has not been experimentally measured.[8] Recent calculations suggest that γ-phase MnSe may be altermagnetic; however, because this phase is unstable in ambient atmospheric conditions, quickly transforming into the more stable α phase, not a lot of characterization has been done.[9,10] Finally, T-phase MnSe behaves similarly to the 2D layered TMD $MnSe_2$ with density functional theory calculations suggesting the existence of high-temperature ferromagnetism ($T_C$ = 250 K) and an indirect band gap of 1.83 eV.[11–13]

Large-scale thin films of MnSe have not yet been realized and we believe this may be due to an insufficient understanding of the growth mechanisms and key parameters controlling lateral crystal growth. In this work, we finally elucidate these mechanisms and parameters by the synthesis of high-quality, low-dimensional MnSe utilizing chemical vapor deposition (CVD). Furthermore, we demonstrate the controllable, repeatable growth of α-phase and β-phase MnSe by CVD and perform a systematic study of the growth parameters impacting large-scale lateral crystal growth. Finally, our magnetic characterization reveals the antiferromagnetic behavior of the β-phase, suggesting that MnSe is a promising material for low-dimensional spintronic devices.

# Results & Discussion

## 1. Growth of α- and β-phase MnSe

The α- and β-phase MnSe crystals were grown on a C-face sapphire substrate by CVD in a three-heat-zone reactor, as shown in Figure 1(a). Both α- and β-phase growths were performed in the same CVD tube using identical precursors, sapphire substrates, and boats. To transition the crystal growth from the α phase to the β phase, a dehydrating step holding all three heat zones of the CVD tube, precursors, and substrate at 250 °C for 35 minutes was introduced prior to the growth process, as illustrated in Figure 1(b) and (c). The low-temperature bake serves two purposes: first, to dehydrate the system and remove any trapped water vapor from the precursors and second, to sublimate the powdered selenium precursor and coat the sapphire substrate with selenium droplets to aid in crystal growth, as confirmed by optical imaging and Raman spectroscopy taken on a sample removed after only the dehydrating bake.

**α-phase Discussion**

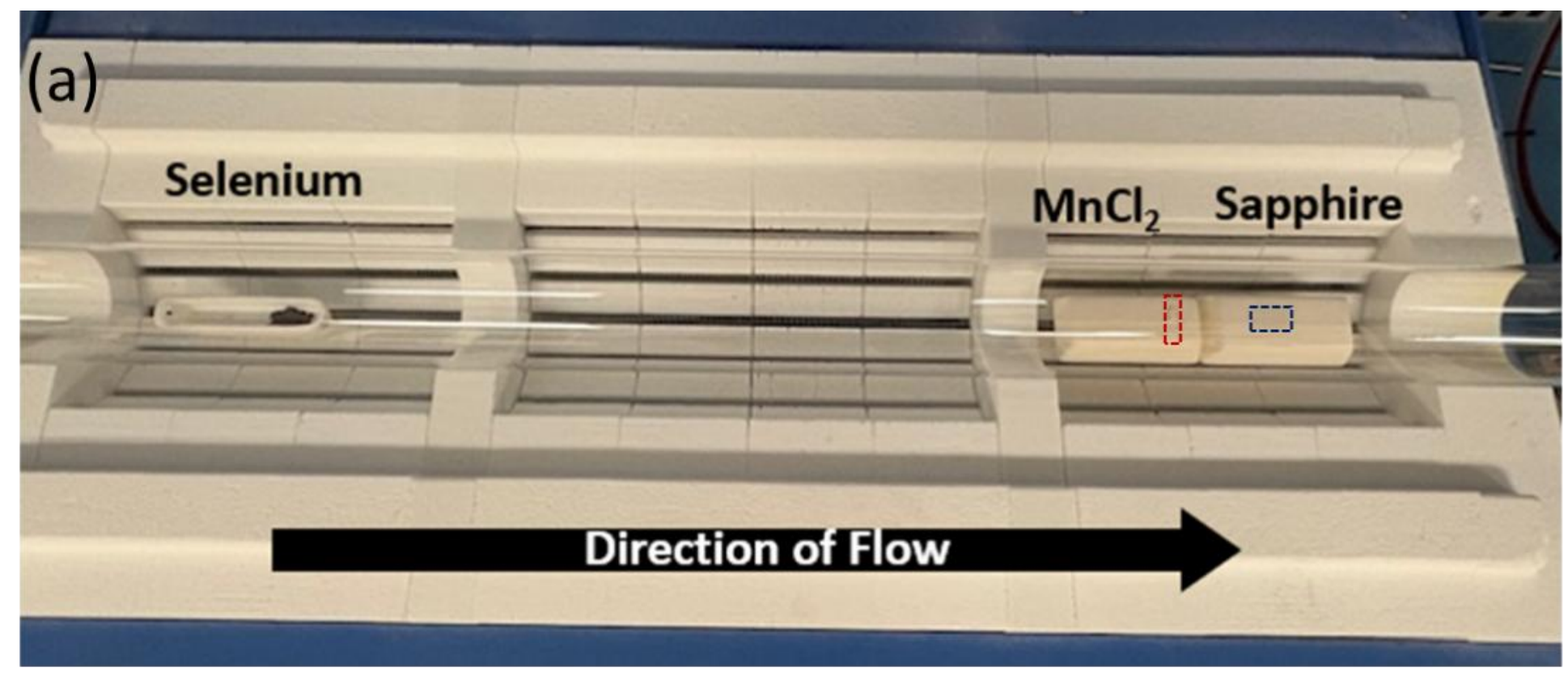

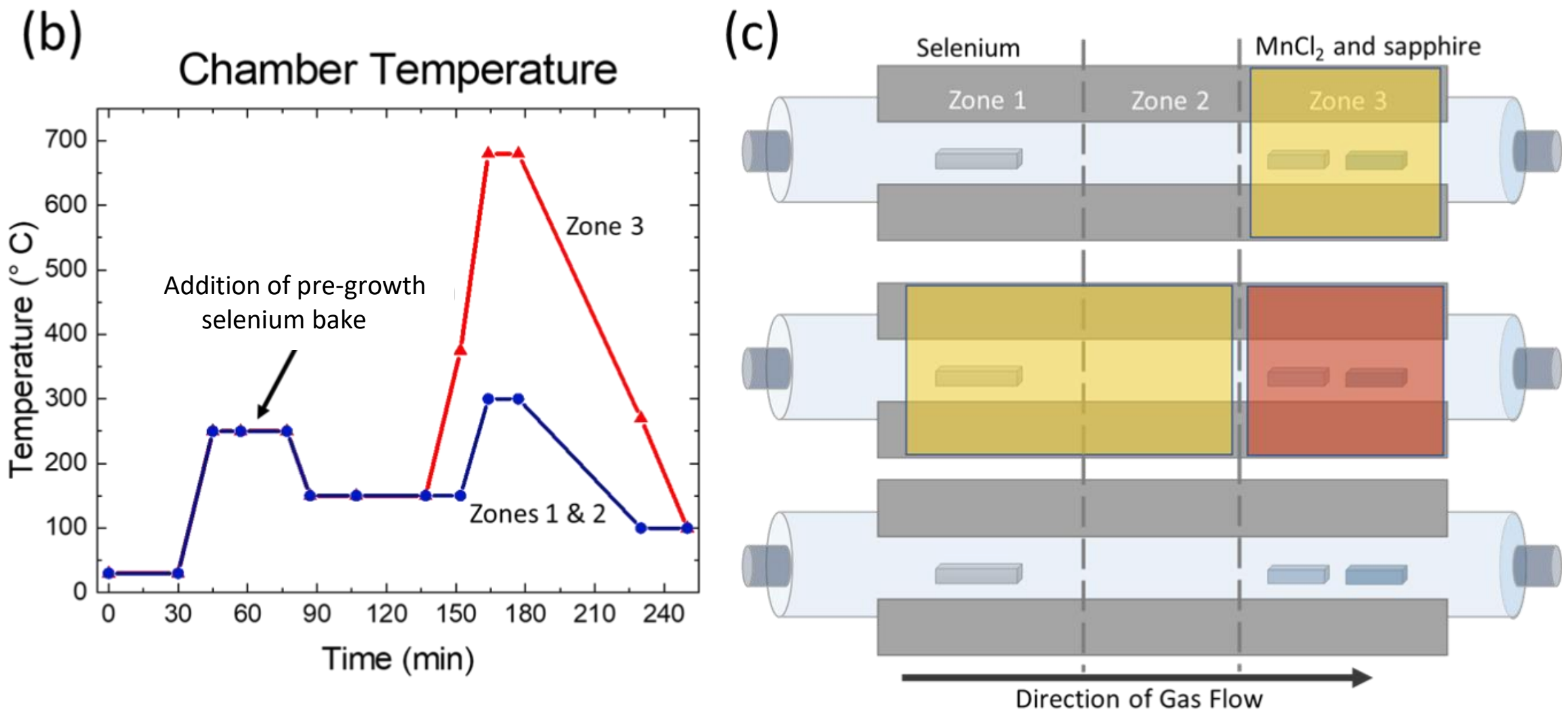

Figure 1. (a) Three-zone chemical vapor deposition furnace with an arrow indicating the direction of gas flow (H/Ar) to the right over the Se, $MnCl_2$ (red rectangle), and C-face sapphire substrates (blue rectangle). (b) Chart of growth recipe showing pre-growth selenium bake and growth steps. (c) Diagram of CVD tube layout with selenium boat in heat zone 1 and $MnCl_2$ and sapphire substrate in heat zone 3 with 3-5 cm separation.

The α-phase MnSe crystals were grown using the CVD recipe shown in Figure 1(a) without the 250°C pre-growth bake described above. Both the precursor configuration and the growth temperature profile are otherwise identical to those in the β-phase recipe. To confirm the crystallographic phase, we performed Raman spectroscopy prior to any detailed morphological analysis. As shown in Figure 2(a), the Raman spectrum exhibits a single dominant peak at 129 $cm^{-1}$, which we assign to the longitudinal acoustic α-MnSe mode, in agreement with prior reports for rock-salt MnSe.[14] We therefore attribute the rod-like crystals in Figure 2(b–e) to the α-phase. Figure 2(b–d) shows that the α-MnSe initially nucleates in "popcorn-kernel-like" clusters that subsequently elongate into rods. The rods preferentially align along three equivalent in-plane directions separated by roughly 60°, consistent with the threefold rotational symmetry of the C-face sapphire surface. A representative AFM line profile, shown in Figure 2(e,f), yields a rod height of ≈ 45 nm and width of ≈ 0.2 μm, with a root-mean-square surface roughness of 240 pm, indicating highly uniform α-MnSe nanorods. The phase uniformity across the α-phase rod in Figure 2(g) indicates a highly ordered, single-domain crystal growth.

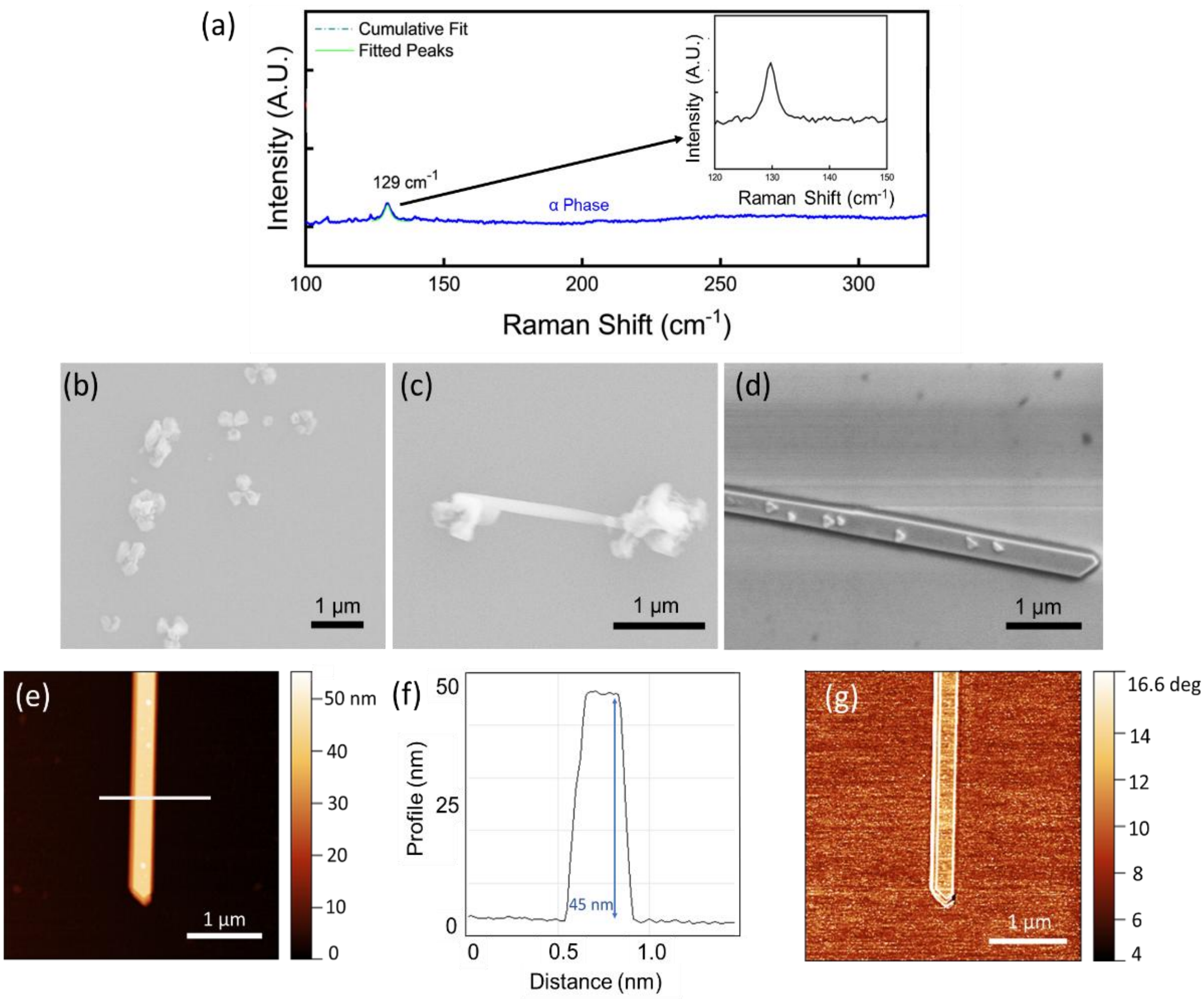

Figure 2. (a) Raman spectra of α-phase MnSe indicating a peak at 129 $cm^{-1}$. (b)-(d) Scanning electron microscope imaging of α-phase MnSe rod nucleation. (e) Atomic force microscopy (AFM) topology and (f) line scan across α-phase MnSe rod of height 45 nm and width 0.2 μm. The white line in (e) indicates the line profile shown in (f). (g) AFM phase mapping of α-phase rod.

**β-phase Discussion**

The β-phase MnSe crystals were grown following the recipe shown in Figure 1(a) with the inclusion of the dehydrating bake prior to the growth steps. Figure 3(a) shows an optical image of the β-phase MnSe flakes in various stages of triangular flake growth, which is further observed with atomic force microscopy (AFM) in Figure 3(b) of a multi-layer β-phase MnSe crystal. The β-phase material grows in both rods and triangular flakes; line profiles of the surface reveal that the growth occurs in the vertical and lateral directions, respectively. The measured flake thickness is 15 nm with additional layers in 5 nm increments.

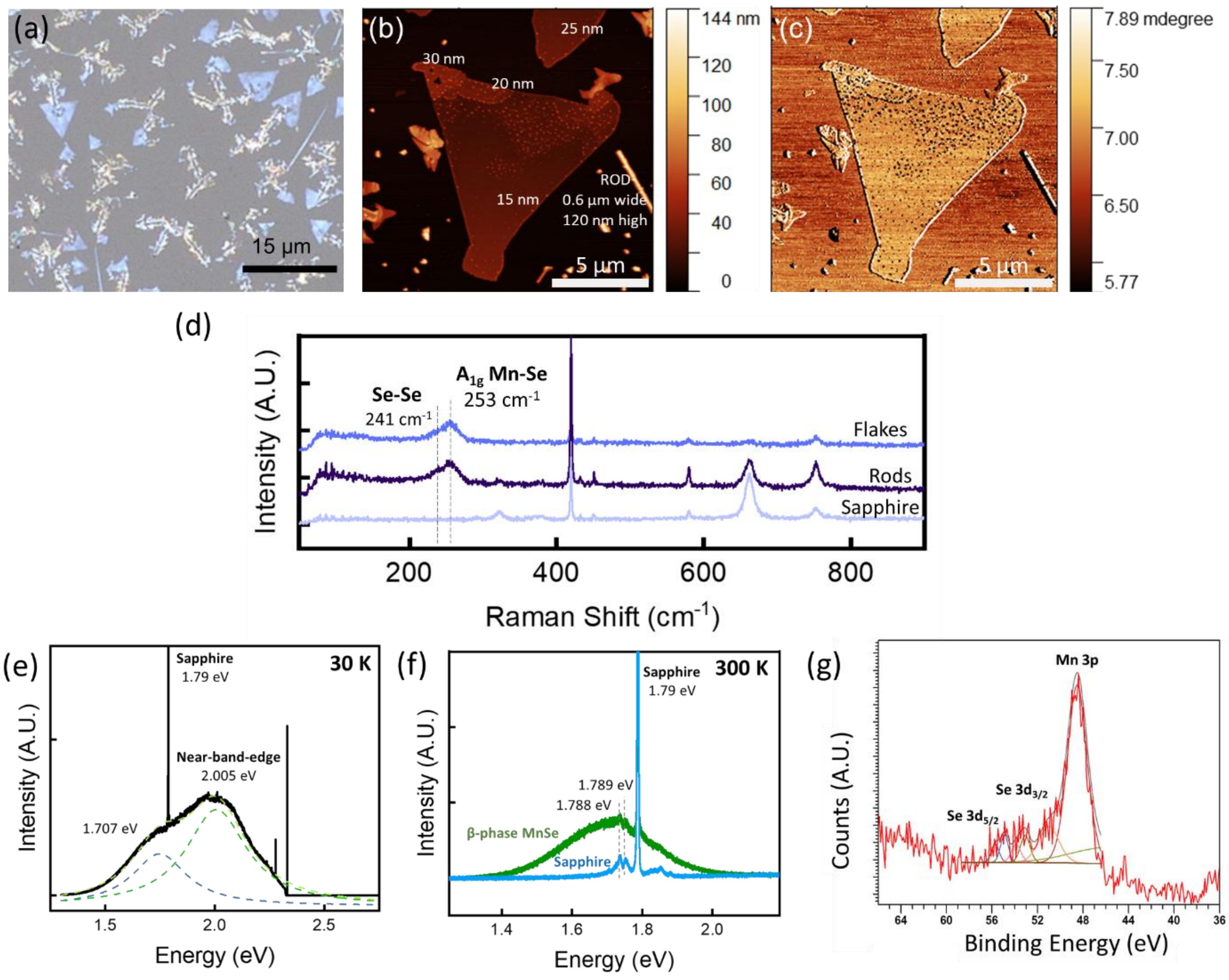

Figure 3. (a) Optical image of β-phase MnSe flake nucleation. (b) Atomic force microscopy and (c) phase imaging of β-phase MnSe flake and rod on a sapphire substrate. (d) Raman spectra of the sapphire substrate and β-phase MnSe flakes and rods. We observe distinct Raman peaks at 241 $cm^{-1}$ and 253 $cm^{-1}$, the Se-Se and $A_{1g}$ Mn-Se stretching modes on the MnSe crystals, respectively. (e) and (f) Photoluminescence spectra of β-phase MnSe on sapphire at 30 K and 300 K, respectively. (g) XPS Se 3d and Mn 3p region spectra of MnSe on sapphire substrate.

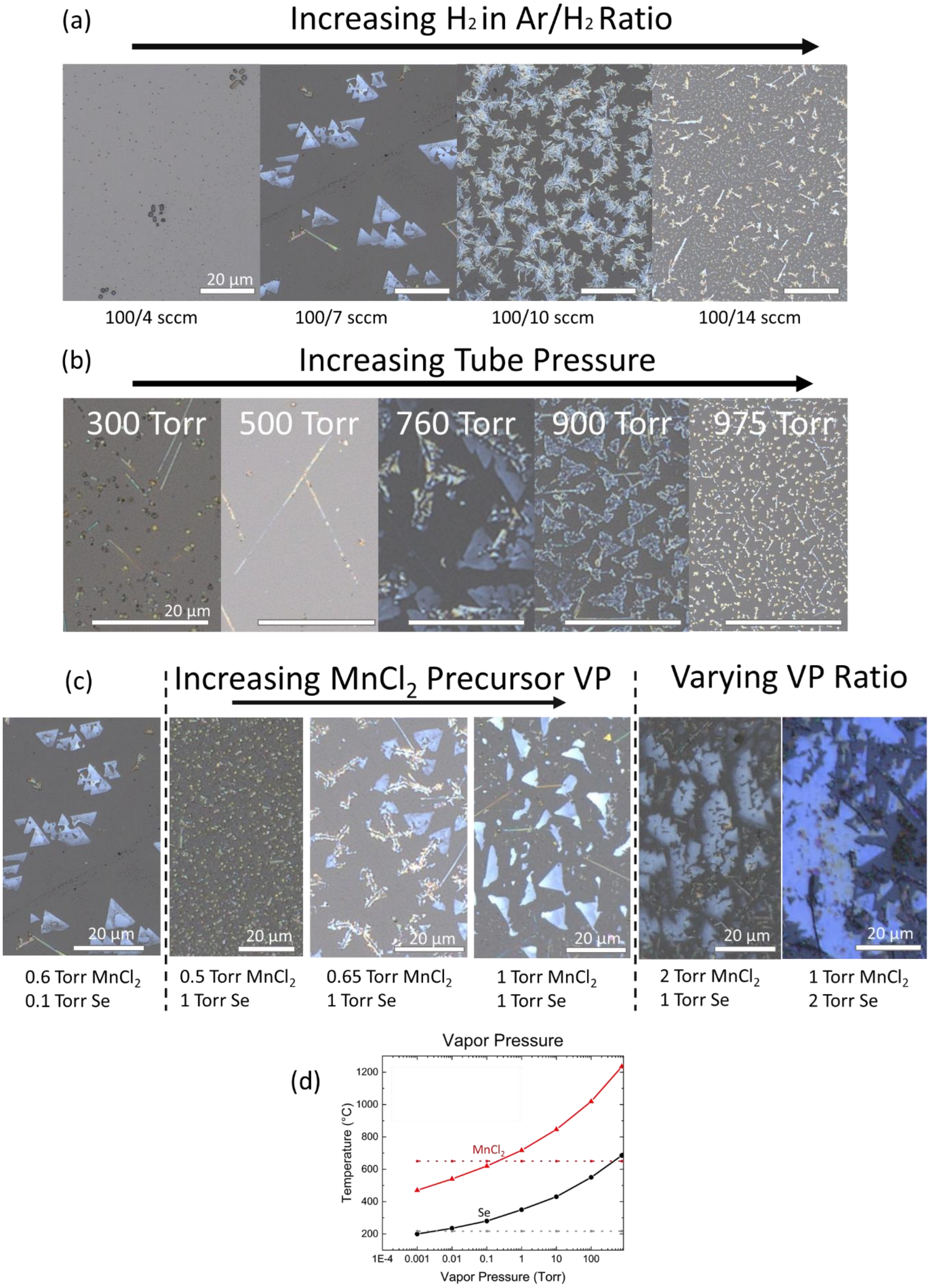

Figure 4. (a)-(c) Impact of Ar/$H_2$ atmospheric concentration, tube pressure, and Se vapor pressure (VP), respectively, on β-phase MnSe crystal growth. (d) Plot of $MnCl_2$ and Se precursor vapor pressures with temperature. Dashed lines indicate the respective precursor melting points. Data taken from reference tables in Stull (1947).

island-then-layer growth mechanism.[15] The AFM phase image in Figure 3(c) shows a uniform phase across the entirety of the flakes, including on the 15 nm to 30 nm layers, suggesting uniform crystal growth. Raman spectra of the β-phase rods and flakes show identical peaks at 241 $cm^{-1}$ and 253 $cm^{-1}$, the Se-Se and $A_{1g}$ Mn-Se stretching modes, respectively,[16] as observed in previous thin-film β-phase MnSe growths.[8] The peak location and FWHM remain consistent for all β-phase morphologies, though the intensity of the 662 $cm^{-1}$ sapphire peak decreases with increased flaked thickness. At low temperatures, the broader MnSe photoluminescence peaks can be distinguished from the sapphire, as shown in Figure 3(e) (30 K). At 300 K (Figure 3(f)), the sapphire peaks[17] at 1.79 eV dominate, but as the peaks broaden with lower temperatures, two distinct curves appear: one centered at 1.707 eV and one at 2.005 eV. As observed with other thin films,[18] the broader peak of the two (E = 1.707 eV) is likely defect induced. The sharper feature centered at 2.005 eV is therefore attributed to the near-band-edge emission of β-MnSe, and we use this value as an estimate of the film band gap, consistent with previous reports on β-MnSe.[8] In the high-resolution Se 3d X-ray photoelectron spectroscopy (XPS), shown in Figure 3(g), the two peaks at 54.8 eV and 55.7 eV are allocated to $3d_{5/2}$ and Se $3d_{3/2}$ for Mn – Se bonding.[19] The location of the Se $3d_{3/2}$ peak at 53 eV instead of at the expected 59 eV suggests that the Se is strongly bound to the Mn atoms within the crystal.[20]

One of the primary challenges with growth of 2D materials is promoting lateral growth over vertical to yield large-scale flakes ideal for device fabrication. While the literature primarily suggests that the hydrogen ($H_2$) partial pressure is responsible for increased lateral growth,[8,21,22] we find that this plays an incidental role in our case, and the greatest factor in promoting lateral growth is in fact the precursor vapor pressure, as shown in Figure 4(a-c). Increasing the $H_2$ in the tube atmosphere (100/4 sccm to 100/14 sccm of Ar/$H_2$, respectively) and tube pressure (300 to 975 Torr) resulted in a variation of crystal morphology, but had no correlation with lateral growth, as shown in Figure 4(a, b), respectively. Increasing the growth time (10 to 30 minutes) increased flake density but had no impact on lateral growth. However, by increasing the vapor pressure of the Se precursor by elevating the temperature in Zone 1 (using the information shown in Figure 4(d)), additional selenium can be sublimated, more readily reacting with the $MnCl_2$ to form crystals up to 20 µm across in size.

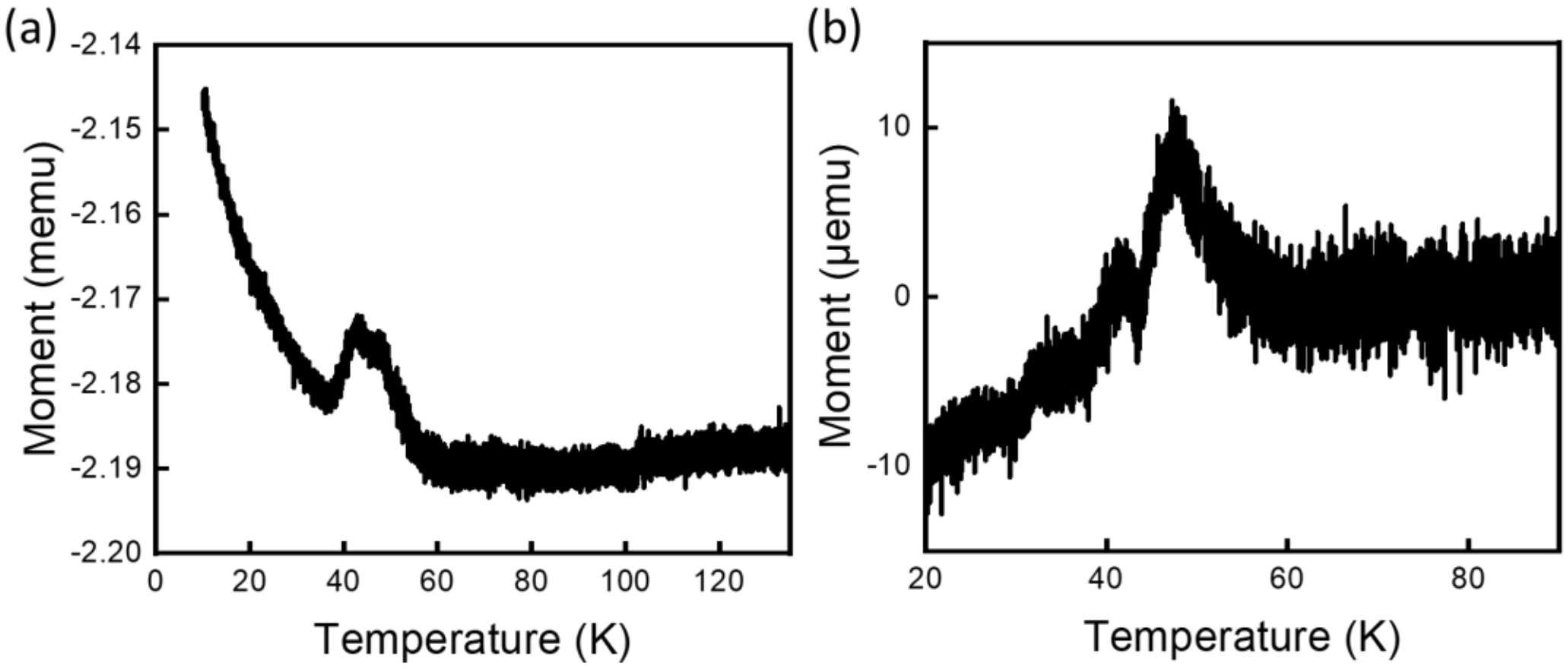

Figure 5. Vibrating sample magnetometry (VSM) of β-phase MnSe on sapphire substrate. (a) Magnetization versus temperature, M(T), of β-phase MnSe on sapphire, showing antiferromagnetic peaks for trapped oxygen and MnSe. (b) Sapphire-background-subtracted M(T) data indicating that β-phase MnSe has a Néel temperature of $T_N$ = 53 K.

Magnetic properties of the β-phase MnSe were investigated using in-plane vibrating-sample magnetometry (VSM). Figure 5(a) shows the as-collected magnetization, M(T), as a function of temperature for a constant in-plane field of 5 T. A broad peak is observed between 40 and 55 K. After subtracting the paramagnetic contribution of the sapphire substrate (Figure 5(b)), the M(T) curve exhibits a single well-defined Néel peak at 53 K, consistent with antiferromagnetic ordering in the β-phase films. The broad peak seen in Figure 5(a) is thus a combination of the Néel peak and that of a peak associated with trapped oxygen in the sapphire.[23] A secondary antiferromagnetic peak is visible at 47 K due to trapped oxygen in the system; however, this peak vanishes in the background-subtracted data. This β-phase antiferromagnetic signature differs from previous magnetization measurements in the literature, which suggested β-phase MnSe is ferromagnetic with a Curie temperature of $T_C$=42.3 K.[8] This disparity is likely due to complex magnetic ordering in the β phase varying with material thickness, as has been observed with layered TMDs.[24] The β-MnSe flakes studied here are ≈ 15–30 nm thick, substantially thicker than the monolayer limit where theory predicts high-temperature ferromagnetism in related MnSe phases.[11,12,25] We therefore interpret our antiferromagnetic signature as characteristic of the multilayer regime. The discrepancy with prior reports of ferromagnetism in β-MnSe ($T_C$ = 42.3 K)[2] may arise from differences in thickness or subtle variations in stoichiometry and strain, but a systematic thickness-dependent study will be required to fully resolve this issue.

In summary, within the ranges of Ar/$H_2$ flow, total pressure, and growth time explored here, the dominant control knob for selecting the MnSe polymorph is the presence or absence of the 250 °C pre-growth bake: growths without the bake reproducibly yield α-MnSe nanorods, whereas growths with the bake reproducibly yield β-MnSe flakes. Other parameters primarily tune lateral flake size and density rather than the phase itself.

## Conclusion

The primary growth phase of MnSe reported in the literature is the cubic α-phase. We have demonstrated the controllable growth of both α- and β-phase MnSe through the addition (or lack thereof) of a 35-minute 250 °C dehydrating bake prior to crystal growth. While the literature primarily suggests that the hydrogen ($H_2$) partial pressure is responsible for increased lateral growth,[8,21,22] we found that this only plays an incidental role in crystal size and the greatest factor in promoting lateral growth is the precursor vapor pressure. Finally, we observed the presence of antiferromagnetism in our β-phase MnSe with a Néel peak at 53 K.

## Experimental Methods

### Growth Methods

MnSe crystals were grown on a C-face sapphire substrate using a three-heat-zone CVD system (PlanarTech) with a two-inch diameter quartz tube. Sapphire substrates are prepared with a 10-minute sonication each in acetone and IPA, and then dried with compressed nitrogen air and annealed in Ar (99.999% purity) at 1000 °C for two hours. Se powder (200 mg, 99.95% purity) is placed in an alumina boat (50 mm x 40 mm x 20 mm) in Zone 1, $MnCl_2$ powder (20 mg, 99% purity) on an upturned boat in Zone 3, and a C-face sapphire substrate placed 4 cm downstream on a second upturned boat in Zone 3; see Figure 1 for more details. The growth is performed with a two-step process: first, a dehydrating 250 °C pre-growth bake with all three heat zones set to the same temperature (250 °C) in an Ar atmosphere (100 sccm) at 760 Torr removes any water vapor from the precursor materials and activates the Se powder by distributing it across the tube. Then, we simultaneously heat Zone 1 and 2 to 300 °C and Zone 3 to 680 °C at a pressure of 250 – 925 Torr and introduce a flow of 100 – 200 sccm Ar and 3-14 sccm $H_2$ into the tube for a growth period of 10-30 minutes. The furnace is then cooled to room temperature over 30 mins and the tube is flushed with 1000 sccm Ar. During this 250 °C pre-growth step, Raman spectra collected from samples removed immediately after the bake show only selenium features, confirming the formation of a discontinuous film of Se droplets on the sapphire surface. These droplets serve as a reactive Se reservoir for subsequent MnSe crystal growth; no residual elemental Se signatures are observed in the fully grown samples. Throughout this work we use the same precursor masses, furnace setpoints, Ar/$H_2$ flow ranges, and growth times for both α- and β-phase growths; the difference between the two recipes is the presence or absence of the 35-minute, 250 °C pre-growth bake in Ar. We therefore refer to the recipe without the pre-growth bake as the "α-MnSe" growth condition and the otherwise-identical recipe with the pre-growth bake as the "β-MnSe" growth condition.

**Material Characterization Methods**

Characterization of the unprocessed α- and β-phase MnSe crystals was performed optically (Keyence Confocal Microscope) and surface morphology was performed with atomic force microscopy (Bruker Dimension FastScan AFM) and scanning electron microscopy (Zeiss Gemini 300) with an energy dispersive spectroscopy detector. Raman and photoluminescence spectra were collected with a Horiba LabRam confocal system utilizing 532 nm excitation through a 100X objective; incident power was 10 mW. X-ray photoelectron spectroscopy (XPS) was performed using a Kratos Axis 165 X-ray photoelectron spectrometer operating in hybrid mode using monochromatized Al $K_\alpha$ radiation ($\hbar\omega$ = 1486.7 eV) at an anode power of 280 W. The survey spectra and high-resolution spectra were collected at pass energies of 160 eV and 40 eV, respectively, under a base pressure of $5\times10^{-8}$ Torr. The angle-resolved XPS measurements were taken at angles of 0°, 40°, 55°, 63°, and 70° with respect to normal incidence. The effect of surface charging on the resulting binding energies was eliminated by shifting the advantageous carbon peak in the C1s to 284.9 eV. Vibrating sample magnetometry (VSM) measurements were performed with a PPMS (Quantum Design).

## Acknowledgements

Funding for the XPS shared facility used in this research was provided by the NSF under the award CHE-1626288.